\documentclass[aps,prl,twocolumn,superscriptaddress,showpacs]{revtex4}
\usepackage{graphicx}
\usepackage{natbib}
\begin{document}

\title{Length scales coupling for nonlinear dynamical problems
  in magnetism}

\author{V.\ V.\ Dobrovitski}
\affiliation{Ames Laboratory, Iowa State University, Ames, Iowa, 50011}
\author{M.\ I.\ Katsnelson}
\affiliation{Ames Laboratory, Iowa State University, Ames, Iowa, 50011}
\affiliation{Institute of Metal Physics, Ekaterinburg 620219, Russia}
\author{B.\ N.\ Harmon}
\affiliation{Ames Laboratory, Iowa State University, Ames, Iowa, 50011}

\date{\today}

\begin{abstract}
The dynamics of real magnets is often governed by several
interacting processes taking place simultaneously at different
length scales. For dynamical simulations the relevant length
scales should be coupled, and the energy transfer accurately
described. We show that in this case the micromagnetic
theory is not always reliable.
We present a coarse-graining approach applicable to
nonlinear problems, which provides a
unified description of
all relevant length scales, allowing a smooth, seamless coupling. 
The simulations 
performed on model systems show that the 
coarse-graining approach achieves nearly the precision of 
all-atom simulations.
\end{abstract}


\maketitle

The dynamics of magnetization in real magnets is often governed
by defects: impurities, vacancies, grain boundaries, etc. They
serve as nucleation centers for magnetization reversal or as
pinning centers for domain walls (DW), due to very localized
(several lattice constants) but significant changes in the
magnetic interactions. As a result, the torques acting on
individual atomic spins in the vicinity of the defect are
strongly non-uniform at the atomic scale. These small-scale
inhomogenieties can result, however, in large-scale (hundreds or
thousands of angstroms) changes of the magnetization distribution
(e.g.\ nucleation of a domain). At larger length scales
(microns), the magnons propagate into the bulk of the sample
carrying away significant energy. Such multiscale phenomena are
encountered in the depinning of DW
\cite{torredw,taras,kron,schafer}, the influence of surfaces
and interfaces on the spins in the bulk 
\cite{kodama,jamet,amils,deak}, etc. These phenomena determine
basic magnetic properties
such as the coercive field and the dynamics of magnetization
reversal.

Accurate simulations of such phenomena might be possible if all
the interactions and dynamics could be treated at the atomic
scale \cite{landau}, but for a sample of the size of few tens of
microns, the number of spins becomes much too large for modern
supercomputers. Atomistic modeling can be used, however, for the
regions close to the defects. For the large-scale processes
occuring far from the defects a collective description should be
used, e.g.\ micromagnetic (MM) theory. To simulate accurately the
multiscale processes, all the relevant scales should be modeled
simultaneously, and seamlessly coupled to describe correctly the
energy transfer. The simplest solution would be to use the
standard MM approach with gradually increasing resolution
(reducing the size of the computational cell) near the defect,
until every micromagnetic cell contains a single atomic spin. At
this ultimate resolution, the micromagnetic 
Landau-Lifshitz-Gilbert equation coincides with the equation of
precession of an individual atomic spin.

In this paper, using simulations performed on model 1-D chain
systems, we show that such an approach is often inadequate for
length scale coupling, giving results which differ drastically
from the exact solutions obtained by all-atom simulations. Thus,
for an important class of dynamical problems in magnetism, a new
approach is needed. Here, we present a theoretical framework
along with a computational scheme, which allows treatment of all
length scales in exactly the same manner, enabling a truly
seamless linking of different scales. The proposed approach is
applicable to essentially nonlinear problems, and yields reliable
results which are in excellent agreement with exact atomic
simulations.

Length scale coupling has been extensively
discussed in the context of lattice dynamics modeling
\cite{kohlhoff,philips,rudd,cai} and successfully applied for the
study of crack propagation and nanoindentation
\cite{rudd1,smith,elefteri}. Analogous schemes can be constructed
for linear magnetic problems \cite{dkh}, but most of the
interesting problems in magnetism are essentially
nonlinear due to the constraint $S_x^2 + S_y^2 + S_z^2 = S^2$
($S_{x,y,z}$ are the components of the spin and $S$ is its
length; everywhere below we assume $S=1$). 
Moreover, the lattice constant $a$ is the relevant
length scale for a typical problem of lattice dynamics, while
magnetic problems have one more intrinsic length
scale, the DW width $\Delta$. Therefore, magnetic
multiscale simulations present some unique problems.

Our approach to linking length scales in dynamical
magnetic simulations is based on statistical
coarse graining (CG), which has been successfully used before
\cite{rudd,dkh} to define suitable collective variables
(``gross variables''). For simplicity, we consider only
the case of zero temperature, thus restricting the analysis
to purely dynamical problems, with dissipation and 
thermal noise effects absent. 
We assume that the system under consideration is 
a ferromagnet made of identical classical spins
${\bf S}_{\mu}$ ($|{\bf S}_{\mu}|=1$), located 
at the $\mu$-th site of the crystalline lattice (greek indices
enumerate the atomic lattice sites). The system is 
described by a rather general spin Hamiltonian
\begin{equation}
\label{ham}
{\cal H} = {\cal H}^0 + {\cal V},\quad
{\cal H}^0 = \sum_{\mu} {\cal H}_{\mu}^0 = \sum_{\mu,\nu} J_{\mu\nu} 
  {\bf S}_{\mu} {\bf S}_{\nu},
\end{equation}
where ${\cal H}^0$ describes the isotropic exchange interaction,
while ${\cal V} \ll {\cal H}^0$ represents
all the other, much weaker interactions, 
and $J_{\mu\nu}$ is the exchange interaction between the 
atomic spins at the sites $\mu$ and $\nu$. 
We will need a Hamiltonian formulation of the 
magnetization dynamics, therefore below we describe spin by
conjugate canonical variables 
$\alpha_{1\mu}=2 \sin{(\theta_{\mu}/2)} \cos{\phi_{\mu}}$ 
and 
$\alpha_{2\mu}=2 \sin{(\theta_{\mu}/2)} \sin{\phi_{\mu}}$,
where $\theta_\mu$ and $\phi_\mu$ are
the azimuthal and the polar angles of the spin vector
correspondingly.
We assume that for the simulations of the large-scale 
regions located far from the defects, the finite-element method
(FEM) is used. Thus, we define the gross variables,
which describe the state of the system at a given length
scale, by averaging the atomic degrees of freedom:
$
\alpha_{1j} = \sum_{\mu} f_{\mu,j}\,\alpha_{1\mu},
$ 
and
$
\alpha_{2j} = \sum_{\mu} f_{\mu,j}\,\alpha_{2\mu},
$
where the index $j$ corresponds already to the computational FEM
nodes. The weight function $f_{\mu,j}$ localized near the node
$j$ must satisfy the normalization condition 
$\sum_{\mu} f_{\mu,j} = 1$. The choice of this function
was discussed \cite{dkh} for linear magnetic multiscale 
problems; our experience shows that the following 
piecewise-linear choice
is sufficient:
$f_{\mu,j} = f_0|\mu-\mu_j|/|\mu_{j}-\mu_{j-1}|$ for 
$\mu\in[\mu_{j-1},\mu_j]$, where $\mu_j$ is the atomic position
of the $j$-th computational node and $f_0$ is the normalization. 

The main assumption of the approaches based on the coarse
graining ideology (such as the non-equilibrium statistical
operator \cite{nso} or projection operator \cite{mori}
methods) is that the atomic degrees of freedom are
in local equilibrium, and the equilibrium conditions are
determined by the gross variables \cite{chant}. This means 
that all atomic spins located near the node $j$ 
are nearly parallel to the direction defined
by the the variables $\alpha_{1,2j}$; the exchange part
${\cal H}_0$ of the Hamiltonian (\ref{ham}) is responsible
for that (thus, the coarse-graining
approach can be used also for finite temperature if it is small
in comparison with the Curie temperature). Then, the distribution 
function for the atomic variables is
\begin{eqnarray}
\label{dist}
\rho &=& Q^{-1} \exp{(-\beta {\cal H}')},\\
  \nonumber
{\cal H}' &=& {\cal H}^0 + \sum_{\mu,j} F_j f_{\mu,j}\,\alpha_{1\mu}
  + \sum_{\mu,j} G_j f_{\mu,j}\,\alpha_{2\mu},
\end{eqnarray}
where $Q$ is the statistical integral, and the  
torques $F_j$ and $G_j$ are determined from the conditions of 
local equilibrium:
$
\sum_\mu f_{\mu,j}\,\langle\alpha_{1\mu}\rangle = \alpha_{1j},
$
and
$
\sum_\mu f_{\mu,j}\,\langle\alpha_{2\mu}\rangle = \alpha_{2j},
$
where the angular brackets mean the statistical 
averaging with the distribution function (\ref{dist}).

However, the exchange Hamiltonian
${\cal H}_0$ is rotationally invariant, and a zero-frequency 
mode is present in the statistical integral $Q$ 
in Eq.\ \ref{dist}.
In the linear approximation, this leads to a divergence,
and the account of nonlinear effects becomes
possible. Such a situation has been studied e.g.\
by Bogoliubov in the work on the small polaron
problem, see e.g.\ Ref.\ \cite{polaron}. The general
ideas of that work can be applied here.

We need to separate the (nonlinear) motion of gross variables from
the (quasilinear) motion of the atomic-scale variables
to exclude the secular terms causing divergence of $Q$.
To do this, we note that
the characteritic scale of nonlinear magnetic structures,
the DW width $\Delta$, is much bigger than the interatomic distance $a$,
and the separation of the slow and the fast 
variables can be done in the
spirit of the modified adiabatic theory \cite{polaron}.
Instead of using the parameters $\alpha_{1,2j}$, we 
define a local coordinate frame, rotated with respect to the
"laboratory" frame by the angles
$\theta_j$ and $\phi_j$ such that 
$\alpha_{1j} = 2 \sin{(\theta_j/2)} \cos{\phi_j}$ and
$\alpha_{2j} = 2 \sin{(\theta_j/2)} \sin{\phi_j}$.
To exclude the secular terms, we use the condition 
$
\sum_\mu f_{\mu,j}\,\langle\alpha^{(j)}_{1\mu}\rangle = 
\sum_\mu f_{\mu,j}\,\langle\alpha^{(j)}_{2\mu}\rangle = 0,
$
where $\alpha^{(j)}_{1,2\mu}$ are defined already 
in
the local coordinate frame associated with $j$-th node.
These conditions are local, since 
the weighting function $f_{\mu,j}$ is local, and the 
statistical integral $Q$ is also calculated in the 
local coordinate frame. As a result, 
the equations of motion for the gross variables
$\theta_j$ and $\phi_j$ can be obtained in closed form:
\begin{eqnarray}
 \nonumber
\dot\theta_j &=& -\sum_{\mu} f_{\mu,j} \left\langle
  {\partial {\cal V}}/{\partial\alpha_{2,\mu}} \right\rangle
  +\sum_{kn} M_{jk} D_{kn} \alpha^{(j)}_{2n}, \\
\label{eqmot}
\sin\theta_j\,\dot\phi_j &=& \sum_{\mu} f_{\mu,j} \left\langle
  {\partial {\cal V}}/{\partial\alpha_{1,\mu}} \right\rangle
  -\sum_{kn} M_{jk} D_{kn} \alpha^{(j)}_{1n},
\end{eqnarray}
where $M_{jk}=\sum_\mu f_{\mu,j}f_{\mu,k}$, and
$D_{jk} = D^0_{jk} - d_jS_k/S_0
- d_kS_j/S_0 + d_0S_jS_k/S_0^2$,
where $D^0_{jk}=\left(\sum_{\mu\nu}f_{\mu,j} {\cal J}^{-1}_{\mu\nu}
f_{\nu,k}\right)^{-1}$, ${\cal J}_{\mu\nu}$ is the Hessian matrix of 
the exchange Hamiltonian, $S_k$ is the eigenvector of 
$D^0_{jk}$ corresponding to the zero eigenvalue, and
$d_j=\sum_l D^0_{jl}$, $d_0=\sum_l d_l$, $S_0=\sum_l S_l$.
Note that the matrices $M$ and $D$ should be calculated only
once and used later during simulations. 

Now, let us focus on the numerical results obtained using the
procedure described above. The first case tests the ability of
our CG scheme to handle the dynamics of an essentially nonlinear
object, e.g.\ a domain wall. We consider a
1-D chain of $N=465$ spins with ferromagnetic nearest-neighbor
coupling $J=25$, and antiferromagnetic next-nearest-neighbor
coupling $J'=-\gamma J= -2.5$. The spins possess single-ion
anisotropy of easy-axis type, $K=0.01$ (the easy axis is
directed along the $y$-direction); the dipole-dipole
interactions are neglected for simplicity. The ends of the chain
are different from the bulk; this represents e.g.\ a nanowire
which has the growth defects at the ends causing the spin
Hamiltonian parameters $J$, $J'$ and $K$ to be different from
their bulk values. Thus, six spins at one end of the chain 
are chosen to have the following parameters:
$J_{01}=0.5 J$, $J_{12}=0.6 J$, $J_{23}=0.7 J$, $J_{34}=0.8 J$,
$J_{45}=J_{56}=J$, $K_0=-0.4 K$, $K_1=-0.2 K$, $K_2=0$, $K_3=0.2 K$,
$K_4=0.4 K$, $K_5=0.8 K$. At the other end of the chain, 
the last six spins have the same parameters
in reverse order. Initially, the chain contains a
Neel DW, with spins rotating in the $x$--$y$ plane from 
the direction
along the $y$-axis to the opposite direction. At $t=0$,
the external field $H=0.01$ is applied in the $x$--$y$ plane, at the
angle $\phi=-0.4\pi$ with the $x$-axis. As a result, the wall starts
moving. After hitting one end of the chain, it deforms, and
reflects, moving to the other end of the chain.
Note that, since we omit dissipation and fluctuations, the
energy of the system is conserved, and the dynamics of the system
is quasi-periodic.

\begin{figure}
\includegraphics[width=6cm]{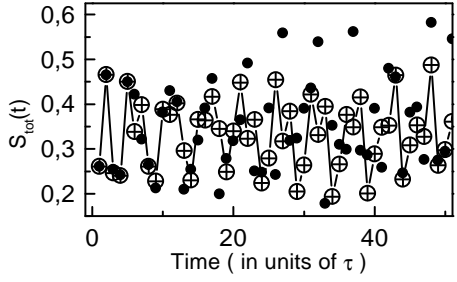}
\caption{\label{waltot} Time dependence of the total magnetization
$S_{\text{tot}}(t)$
of the chain calculated by three methods: large white circles ---
atomic simulations, small black circles --- MM simulations,
crosses --- CG scheme. Initial conditions correspond to the
Neel domain wall. Since the CG results almost coincide with
the atomistic ones, the crosses appear mostly in the centers
of the white circles.}
\end{figure}

\begin{figure}
\includegraphics[width=6cm]{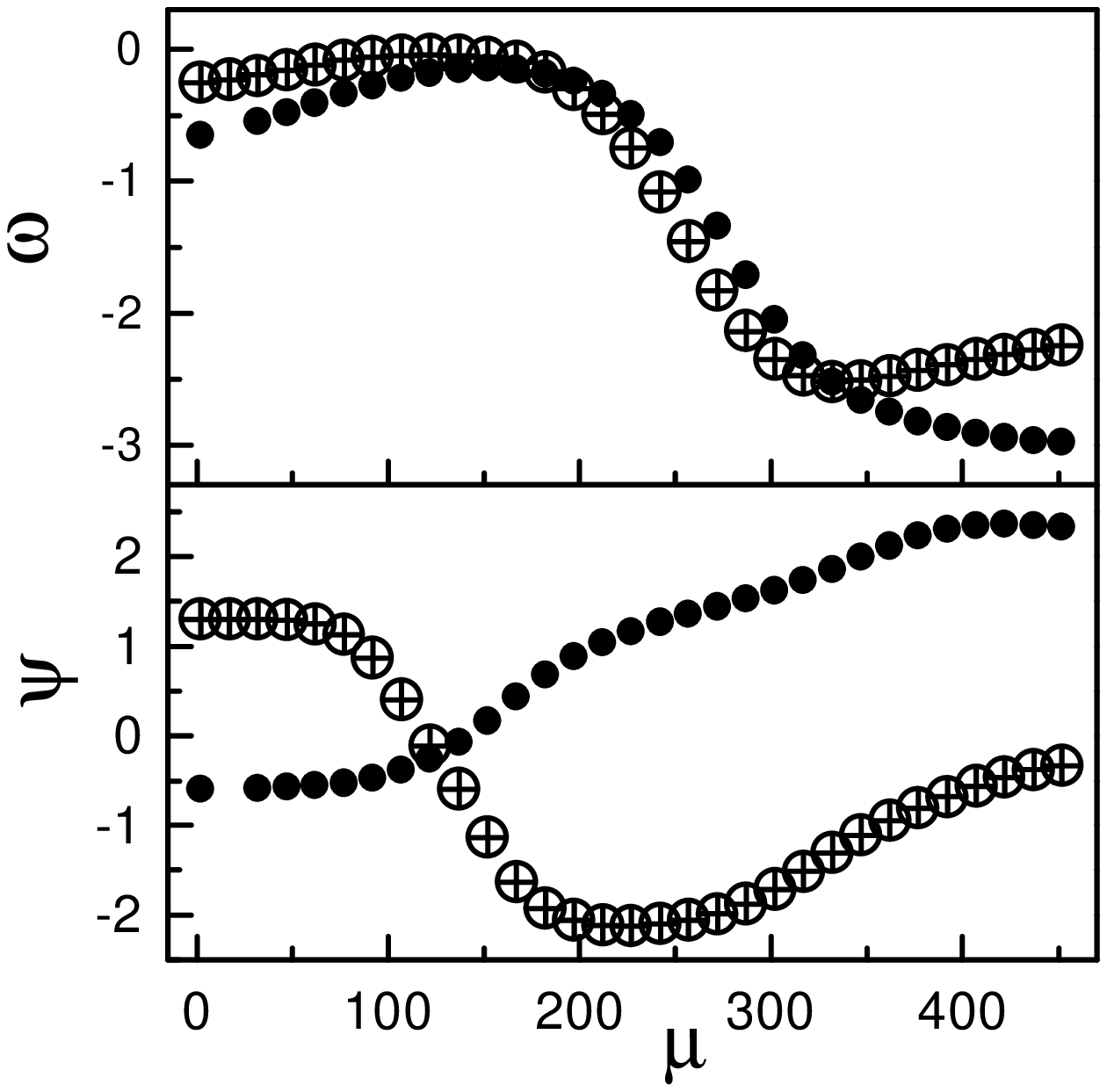}
\caption{\label{walprof} The magnetization profile
in the chain 
(the angles $\omega_\mu$ and $\psi_\mu$)
at $t=60\tau$ calculated by three methods: 
large white circles ---
atomic simulations, small black circles --- MM simulations,
crosses --- CG scheme. Initial conditions correspond to the
Neel domain wall. Since the CG results almost coincide with
the atomistic ones, the crosses appear mostly in the centers
of the white circles. The values of $\omega_\mu$ and $\psi_\mu$
at the nodes of the MM grid are shown.}
\end{figure}

To compare the performance of different computational schemes, 
we have modeled the system's dynamics by (1) atomic simulations
which give the exact solution, (2) by micromagnetic simulations
using the method suggested in Ref.\ 
\cite{fredkin}, with the number of spins in the cell varying
from five (in the middle of the chain) to one (at the ends),
and (3) by the CG scheme described above,
with
the same grid as used in the MM simulations.
The computational time step, along with all other relevant
parameters, has been kept the same for all schemes.
The temporal dependence of the chain's total magnetization
$S_{\text{tot}}(t)=\sqrt{S^2_x(t)+S^2_y(t)+S^2_z(t)}$
is shown in Fig.\ \ref{waltot},
where the time is measured in units of 
$\tau=0.4\pi/|H\sin{\phi}|\approx 132$.
As one can see, the MM simulations (black circles) differ 
considerably
from the exact atomistic solution, while the CG scheme gives
results practically coinciding with the atomistic simulations.
The same conclusion can be drawn from Fig.\ \ref{walprof}, where
the magnetization profile in the chain is shown at $t=60\tau$;
the magnetization direction at every atomic site ${\bf S}_\mu$
is characterized by the two angles, $\omega_\mu$ and $\psi_\mu$
such that $S^x_\mu=\sin{\omega_\mu}\cos{\psi_\mu}$, 
$S^y_\mu=\cos{\omega_\mu}$ (since the easy axis coincides with
the $y$-axis), and $S^z_\mu=\sin{\omega_\mu}\sin{\psi_\mu}$.

A similar set of simulations has also been performed to study
the magnetization reversal in the chain.
The system, along with all simulation parameters, has been kept
the same as described above, but the initial condition
represents a chain magnetized to saturation, $S^y_{\mu}=1$ for
all $\mu$. At $t=0$, again, the external field $H=0.02$ has been
applied, at the angle $\phi=-0.4\pi$ to the $x$-axis.

\begin{figure}
\includegraphics[width=6cm]{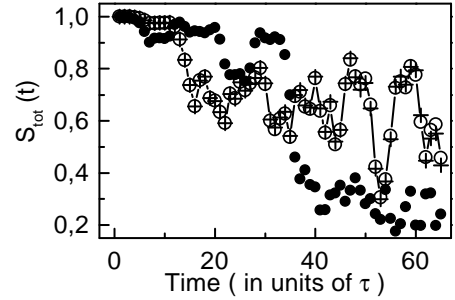}
\caption{\label{revtot} Time dependence of the total magnetization
$S_{\text{tot}}(t)$
of the chain calculated by three methods: large white circles ---
atomic simulations, small black circles --- MM simulations,
crosses --- CG scheme. Initial conditions correspond to chain
uniformly magnetized along the $y$-axis. Since the CG results 
almost coincide with
the atomistic ones, the crosses appear mostly in the centers
of the white circles.}
\end{figure}

Since we are studying an energy-conserving case, in the absence
of the defects at the ends of the chain, all the spins would
rotate in unison. However, in the presence of the perturbation
caused by the defects, different spins rotate with slightly 
different rates, and the system's motion becomes stochastic.
Therefore, after some time, the Zeeman energy is transferred
to the energy of short-wavelength magnons, leading to a 
gradual decrease of the system's total magnetization
\cite{bert,bert1}.

\begin{figure}
\includegraphics[width=6cm]{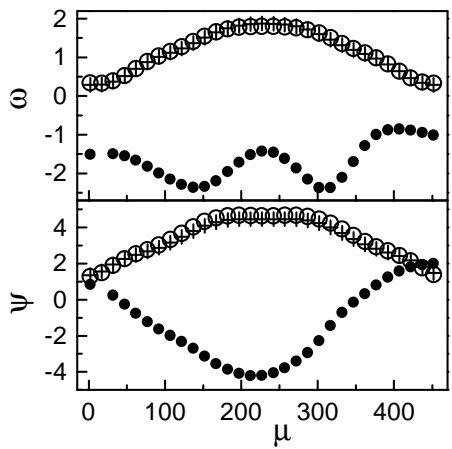}
\caption{\label{revprof} The magnetization profile
in the chain (the angles $\omega_\mu$ and $\psi_\mu$)
at $t=60\tau$
calculated by three methods: 
large white circles ---
atomic simulations, small black circles --- MM simulations,
crosses --- CG scheme. Initial conditions correspond to chain
uniformly magnetized along the $y$-axis.}
\end{figure}

The chain's total magnetization as functions of time is
shown in Fig.\ \ref{revtot}, with the time unit 
$\tau=0.4\pi/|H\sin{\phi}|\approx 66$. Magnetization profiles at 
$t=60\tau$ are presented in Fig.\ \ref{revprof}. Again, one
can see that the MM approach does not describe
the behavior of the system correctly even at a qualitative
level. The CG scheme performs much better, giving the 
results very close to the exact atomistic solution.

Such a drastic difference in the two schemes, 
CG and MM, is caused by
the essential nonlinearity of magnetization dynamics. Due to the
presence of atomic-scale inhomogenieties, in the course
of the system's motion, a considerable number of short-wavelength
excitations are generated near the defect. 
Wavelengths of these magnons are
smaller than the size of the computational cell, and the
micromagnetic theory neglects them completely, although 
the collective effect of a large
number of such modes is significant. In contrast, the
coarse-graining approach does take the atomic-scale modes into
account, and describes very well the collective properties
of an ensemble of these short-wavelength excitations.

In summary, we have shown that the standard micromagnetic theory
does not handle correctly the dynamics of nonlinear multiscale
magnetic processes. We have suggested another approach, based on
statistical coarse graining, which is applicable to nonlinear
problems. Numerical tests on 1-D systems show that the CG scheme,
in contrast with the MM approach, gives almost exact results for
rather large time spans. It is worthwhile to note that the only
essential ingredient of the coarse graining approach is the
standard theory of local equilibrium. Therefore, the CG scheme
can be applied to a large set of 2-D and 3-D problems, including
dissipation, thermal noise, interaction with other degrees of
freedom (spin-phonon interactions) etc.

The authors would like to thank J.\ Morris, R.\ Sabiryanov, and
R.\ Rudd for helpful discussions.
This work was partially carried out at the Ames Laboratory, which 
is operated for the U.\ S.\ Department of Energy by Iowa State 
University under Contract No.\ W-7405-82 and was supported by 
the Director of the Office of Science, Office of Basic Energy Research
of the U.\ S.\ Department of Energy. This work was partially supported 
by Russian Foundation for Basic Research, grant 00-02-16086.


\end{document}